# A Disk Scheduling Algorithm Based on ANT Colony Optimization

Hossein Rahmani
*Electrical and Computer Engineering Department*
*Shahid Beheshti University*
*G.C; Tehran, Iran*
*h.rahmani@mail.sbu.ac.ir*

Sajjad Arshad
*Electrical and Computer Engineering Department*
*Shahid Beheshti University*
*G.C; Tehran, Iran*
*s.arshad@mail.sbu.ac.ir*

Mohsen Ebrahimi Moghaddam
*Electrical and Computer Engineering Department*
*Shahid Beheshti University*
*G.C; Tehran, Iran*
*m_moghadam@sbu.ac.ir*

**Abstract**

Audio, animations and video belong to a class of data known as delay sensitive because they are sensitive to delays in presentation to the users. Also, because of huge data in such items, disk is an important device in managing them. In order to have an acceptable presentation, disk requests deadlines must be met, and a real-time scheduling approach should be used to guarantee the timing requirements for such environment. However, some disk scheduling algorithms have been proposed since now to optimize scheduling real-time disk requests, but improving the results is a challenge yet. In this paper, we propose a new disk scheduling method based on Ant Colony Optimization (ACO) approach. In this approach, ACO models the tasks and finds the best sequence to minimize number of missed tasks and maximize throughput. Experimental results showed that the proposed method worked very well and excelled other related ones in terms of miss ratio and throughput in most cases.

***Keywords:*** *Disk Scheduling, ACO, real time, missed task*

## 1. INTRODUCTION

Processor speed is expected to be doubled every year due to the advances in hardware technology. The capacity of storage devices is also increased at 60% to 80% annually, but no similar improvement is expected to reduce the disk access time [1]. Due to relatively slow speed of disks the role of an efficient disk scheduler algorithm is very crucial to deliver a smooth video playback to the users. On the other hand, one of the requirements of real-time applications is Quality of Service (QOS) guarantee by operating system [2]. These applications are categorized based on the strictness of their QOS requirements as soft or hard real-time applications [3]. In soft real-time applications such as video/audio playback, the most important QOS requirements are minimizing the number of missed deadlines requests while maximizing the system throughput [4,5]. In multimedia soft real time systems, disk scheduling has an important role in satisfying real time constraints.

There are some traditional disk scheduling algorithms such as FCFS, SCAN, C-SCAN, LOOK, C-LOOK and SSTF [6,7,8,9] that do not consider real-time constraints of I/O tasks and therefore are not suitable to be applied directly on a real-time system. On the other hand, some other disk scheduling algorithms such as Earliest-deadline-first (EDF) address this issue without considering disk-seek time as an important bottle neck in the systems. The employment of EDF in the strict sense results in poor throughput and excessive seek-time.

SCAN-EDF [10], which utilizes SCAN to reschedule tasks in a real-time EDF schedule, is one of the best-known real-time disk scheduling algorithms. Since tasks rescheduled in SCAN-EDF should have the same deadline, its efficiency depends on the number of tasks with the same deadlines. If all tasks have different deadlines, the schedule result of SCAN-EDF would be the same as EDF. In SCAN-EDF algorithm, rescheduling is only possible within a local group of requests. To overcome this problem, Deadline- Modification-SCAN (DM-SCAN) [11] suggests the use of Scannable-groups. In this algorithm, request deadlines are reduced several times during the process of rescheduling to preserve EDF schedule. Unlike DM-SCAN, Reschedulable-group-SCAN (RG-SCAN) [12] does not require its input disk requests to be sorted by their deadlines. It also forms larger groups without any deadline modification.

In SCAN-EDF, DM-SCAN, and RG-SCAN algorithms rescheduling is only possible within a local group of requests. Chang et al. in [13] suggests Global Seek-optimizing Real-time (GSR) disk scheduling algorithm that groups the EDF input tasks based on their scan direction. These tasks are moved to their suitable groups to improve the system performance in terms of increased disk throughput and decreased number of missed deadlines. GSR schedules are always feasible if the input real-time disk requests are arranged in an EDF feasible sequence. But with an infeasible input, it is very unlikely to have a feasible output.

The general real-time disk scheduling with linear seek-cost function is an NP-complete problem [14], therefore, ACO (Ant Colony Optimization) may be employed to solve it. To the best of our knowledge, there is only one ACO disk scheduling methods in literature that has been proposed by S. Okdem et al. in [15]. This method works based on the idea of travelling salesman problem (TSP) and aims to reduce the response time of requests. In this method no solution has been considered to reduce the number of missed requests.

Here, we propose an ACO based method to schedule disk requests. The proposed method considers reducing missed requests while tries to improve throughput. Experimental results were satisfactory and showed the proposed method worked better than related ones in terms of miss ratio and performance.

The rest of paper is organized as follows: in section 2, the real-time disk scheduling problem is described briefly and in section 3, the proposed approach is introduced. Section 4 describes the evaluation results and simulation way and in section 5, paper is concluded.

## 2. PROBLEM DESCRIPTION

Each disk request $T_i$ in a real-time environment is defined by its ready time $r_i$, deadline time $d_i$, sector number $l_i$, data size $b_i$, and its corresponding track location $a_i$. Ready time is the earliest time at which a disk task can start. Deadline time is the latest time at which disk task should be completed. The actual starting and completing time of a disk task are called start time $s_i$ and fulfill (finish) time $f_i$ respectively. The start time and finish time of a real-time task $T_i$ with schedule sequence $T_jT_i$ are computed by $s_i = \max\{r_i, f_j\}$ and $f_i = s_i + c_{j,i}$, respectively. The start time and finish time of a real-time task $T_i$ with schedule sequence $T_jT_i$ are computed by $s_i = \max\{r_i, f_j\}$ and $f_i = s_i + c_{j,i}$, respectively.

Assume that the schedule sequence consists of two sequential tasks $T_j$ and $T_i$.To serve the disk request $T_i$, the disk-head moves from previous task cylinder ($a_j$) to the requested one ($a_i$) by a seek-time cost. Then a rotational latency is used for the desired sector. Finally, the requested data ($b_i$) are transferred from disk to buffer in a transfer time. Therefore, the service time of task $T_i$ calculated as follows:

$$c_{j,i} = SeekTime(|a_i - a_j|) + RotationalLatency(l_i) + TransferTime(b_i) \qquad (1)$$

Consider the schedule sequence $t': T_{w(0)}T_{w(1)} \ldots T_{w(i)} \ldots T_{w(n)}$; schedule fulfill time ($f_{w(n)}$) is the finish time of the latest task ($T_{w(n)}$) and $b_{w(i)}$ is the data size of request $T_{w(i)}$. Therefore, the disk throughput is calculated as follows when system start time is zero:

$$Throughput = \sum_{i=1}^{n} b_{w(i)}/f_{w(n)} \propto (f_{w(n)})^{-1} \qquad (2)$$

Therefore, the problem objective that is defined to maximize throughput can be achieved by minimizing the schedule fulfill time while number of missed tasks is minimized. Overall, a real-time disk scheduling problem is defined as follows:

**Definition 1. Consider a set of $n$ real-time disk tasks $t = \{T_0T_1 \ldots T_i \ldots T_n\}$. Finding a schedule with maximal throughput and minimum missed tasks, is the goal of real-time disk schedulers.**

## 3. PROPOSED APPROACH

ACO algorithm was developed by Dorigo et al. [16]. These algorithms evolve their social behavior based on the fact that ants are able to find the shortest route between their nest and a source of food. This is done using pheromone trails, which ants deposit whenever they travel, as a form of indirect communication.

Ant System, as the first ant colony optimization algorithm, showed to be a viable method for attacking hard combinatorial optimization problems. In the proposed approach, we used *MAX-MIN* Ant System (*MMAS*) [17] that is an Ant Colony Optimization algorithm derived from Ant System. The *MMAS* algorithm has made two changes in the standard Ant System in order to optimize its performance:

1. In MMAS, in order to achieve a strong exploitation of the search history, only the best solution of each iteration is allowed to add pheromone during the pheromone trail update.

2. MMAS uses a rather simple mechanism for limiting the strengths of the pheromone trails. A lower bound and an upper bound are enforced for the total amount of pheromone left on each edge. This solution effectively avoids premature convergence of the search.

The proposed ACO based disk scheduling algorithm aims to find the best order of tasks according to two objectives: minimizing the total number of missed tasks and maximizing the disk throughput. Because tasks enter and leave the system dynamically at any moment, this algorithm chooses the best possible task, among the tasks already exist in the queue. The algorithm is run simultaneously with the disk operation, in this way the best next tasks is selected while the current task is being serviced. The pseudo code of proposed method is as follows:

```
Procedure ServiceTasks;
CurrentTime = 0;
CurrentCylinder = 0;
While (queue is not empty) Do
    Remove missed tasks from queue;
    NewTask =
    GetNextTask(queue, CurrentTime, CurrentCylinder);
    Select NewTask to execute;
    CurrentTime+= SeekTime(|a_NewTask − CurrentCylinder|)
                  + RotationalLatency(l_NewTask)
                  + TransferTime(b_NewTask);
    CurrentCylinder = a_NewTask;
    Update queue with tasks satisfy in: ReadyTime of task
    ≤ CurrentTime;
End;
Procedure GetNextTask;
Initialize m = (number of tasks in queue) / 2;
Initialize pheromone of every edge (T_i, T_j) = τ_max;
Initialize Ant parameters;
BestGlobalAnt = null;
While (number of iterations < m and not coverage) Do
```

$Generate\ m\ solutions\ (ants);$
**For** $(each\ ant\ k)$
$Construct\ list\ L_k;$
**For** $(each\ ant\ k)$
$Calculate\ Fitness\ of\ ant\ k;$
$Find\ the\ best\ ant\ for\ current\ iteration\ (BestLocalAnt);$
$Update\ the\ pheromone\ trail\ by\ BestLocalAnt;$
**For** $(each\ edge\ (T_i, T_j))$
**If** $(\tau(T_i, T_j) > \tau_{max})$ **Then** $\tau(T_i, T_j) = \tau_{max};$
**Else If** $(\tau(T_i, T_j) < \tau_{min})$ **Then** $\tau(T_i, T_j) = \tau_{min}$
**If** $(Fit_{BestLocalAnt} < Fit_{BestGlobalAnt})$ **Then**
$BestGlobalAnt = BestLocalAnt;$
**End**;
**Return** $the\ first\ task\ from\ the\ BestGlobalAnt's\ list;$

In the procedure $ServiceTasks$, each time the scheduler wants to pick a task, the queue has to be updated. So all the missed tasks, and also the tasks that will surely be missed by the end of current task are deleted from the queue. As the running time of current task and its cylinder is known, the future values of $CurrentTime$ and $CurrentCylinder$ at the end of current task are calculated, and tasks which will be missed by that time are distinguished and deleted. So each task with $ReadyTime$ less than $CurrentTime$ is added to queue. Then the $GetNextTask$ procedure is called in order to identify the next best request to execute. Ants move on the tasks in order to find the best possible solution. At first, $m$ solutions (ants) are generated while each one has a list $L_k$ which is used to keep the order of tasks for being serviced; this list is emptied at the beginning. Initially, each ant randomly selects one of the tasks in the queue as the first task and adds it to its list. Then in each step the ant selects one of the unselected tasks from queue by considering the selection probability that is defined in Eq. 3. We name the last selected task in the list of an ant as $T_i$ and any other unselected task is named as $T_j$. When ant $k$ wants to select a new task and add it to its list $L_k$, a probability function $P$ is calculated for each $T_j$ according to the following equation:

$$P_k(T_i, T_j) = \begin{cases} \dfrac{\tau(T_i, T_j)^{\alpha} \times \eta(T_i, T_j)^{\beta}}{\sum_{T_j \notin L_k} \tau(T_i, T_j)^{\alpha} \times \eta(T_i, T_j)^{\beta}} & if\ T_j \notin L_k \\ 0 & otherwise \end{cases} \quad (3)$$

$P_k(T_i, T_j)$ is the probability that ant $P$ selects $T_j$ as its next task. Here $\alpha$ and $\beta$ are exponent parameters that control the relative importance of pheromone concentration versus the heuristic factor. Both $\alpha$ and $\beta$ can take values greater than zero and should be determined by trial and error. $\tau(T_i, T_j)$ is the amount of pheromone on the path between $T_i$ and $T_j$, and $\eta(T_i, T_j)$ is a heuristic function that is defined as follows:

$$\eta(T_i, T_j) = \frac{1}{\gamma \times SeekTime(|a_i - a_j|) + (1-\gamma) \times (d_j - f_i)} \quad (4)$$

As our objectives are both minimizing the number of missed tasks and maximizing the disk throughput, we consider a variable $\gamma$ which defines the effectiveness of these two parameters on our heuristic function. By decreasing $\gamma$, the effectiveness of deadline is increased and hence the miss count would be degraded. On the other hand, increasing $\gamma$, results in throughput increase.

Before an ant selects a new task from the queue, it virtually simulates the system's state after choosing that task. This is done by calculating the $CurrentTime$ and $CurrentCylinder$ of system in that situation. Consequently, during this simulation some of the tasks may be missed and deleted from the virtual queue of the ants.

After the virtual queue of ant $k$ is emptied, the ant's job is over and its result sequence will be evaluated by the following fitness function:

$$Fit_k = MakeSpan + MissCount \times WorstDeadline \quad (5)$$

Where $MakeSpan$ is the total execution time of all tasks in the list, $MissCount$ is the number of missed tasks, and $WorstDeadline$ is the maximum deadline of tasks in the queue. As Eq. 5 shows, the fitness function is minimum when the value of $MissCount$ is zero and the $MakeSpan$ has its minimum value. By using this function, when $MissCount$ is minimized, number of completed task is maximized, therefore, the throughput increases. On the other hand, by minimizing the $MakeSpan$, the throughput increases also. Therefore, the proposed fitness function, models increasing throughput, decreasing miss ratio and decreasing $MakeSpan$ simultaneously.

After all ants have prepared their result sequences, the pheromone of each path $(T_i, T_j)$ is modified by the following formula:

$$\tau^t(T_i, T_j) = \rho\tau^{t-1}(T_i, T_j) + \Delta\tau^t(T_i, T_j) \quad (6)$$

$$\Delta\tau^t(T_i, T_j) = \begin{cases} \frac{1}{Fit_{best}} & if\ (T_i, T_j)\ \epsilon\ BestLocalAnt's\ path \\ 0 & otherwise \end{cases} \quad (7)$$

Where $\tau^t(T_i, T_j)$ is the amount of pheromone on path $(T_i, T_j)$ in the current iteration, $\tau^{t-1}(T_i, T_j)$ has the same value with previous iteration, $\rho$ is the evaporation parameter, and $Fit_{best}$ is the fitness of best ant's result in the current iteration.

After termination, the first task of the best global ant's result is returned for execution. The termination condition satisfies when number of iteration is greater than half of tasks or coverage occurs.

## 4. EXPERIMENTAL RESULTS

In this section, experimental results are presented that consists of the comparison results among the proposed method and some other related ones. All implementations performed on a personal computer with 1.66 GHZ of CPU and 2 GB of RAM in the C++ environment. In getting results a typical disc (HP 97560) that its main parameters are shown in Table 1.

Test input consists of a collection of disk requests with their ready times assigned automatically by a uniform random number. The deadline is relatively calculated by summation of the corresponding ready time with the period time of task. Period time is also distributed uniformly. Each task has a 36*KB* request for data. The test sets used in the simulations are shown in Table 2.

Table 1- Disk parameters of HP 97560

| | |
|---|---|
| Cylinders per disk | 1972 |
| Tracks per cylinder | 19 |
| Sectors per track | 72 |
| Sector size | 512 bytes |
| Seek time function (ms) | $\text{Seek}(D_{j,i})=\begin{cases}3.24+0.4\sqrt{D_{j,i}} & D_{j,i}\le 383\\ 8.00+0.008D_{j,i} & D_{j,i}>383\end{cases}$ |
| Revolution speed | 4002 *RPM* |
| Transfer time | 10 *MBps* |

Table 2- Test sets used in simulation

| Test sets | Number of Problem | Ready time | Number of Tasks |
|---|---|---|---|
| TC1 | 1000 | Uniform Random(0..160*ms*) | 20 |
| TC2 | 1000 | Uniform Random(0..240*ms*) | 30 |
| TC3 | 1000 | Uniform Random(0..400*ms*) | 50 |

The results are compared with some well-known methods such as C-LOOK, CSCAN, FIFO, EDF, SCAN-EDF, GSR and Proposed ACO in terms of number of missed tasks and achieved throughput.

It is worth mentioning that the parameters of ACO were adjusted by the values that are shown in Table 3.

Table 3- Parameters of ACO (N: Number of tasks)

| Number of ANTs | Maximum Iterations | $\alpha$ | $\beta$ | $\rho$ | $\gamma$ | $\tau_{min}$ | $\tau_{max}$ |
|---|---|---|---|---|---|---|---|
| $\frac{N}{2}$ | $\frac{N}{2}$ | 1.0 | 2.0 | 0.98 | 0.1 | 10 | 20 |

Figure 1 and 2 shows the average miss ratio and throughput for all algorithms when they were applied on TC1 problems. For each problem we applied the proposed ACO, 100 times the average number of missed tasks and throughputs have been reported.

It is obvious in the figure 1 that the proposed ACO has better results in terms of the miss ratio in compare with other methods. The proposed ACO has improved the miss ratio about 7% in the average case in compare with GSR which has the lowest miss ratio among related methods. Also, our method worked better than FIFO, EDF, GSR and SCAN-EDF in terms of throughput. Its throughput is 5.27*36*KBps* more than GSR throughput in the average case.

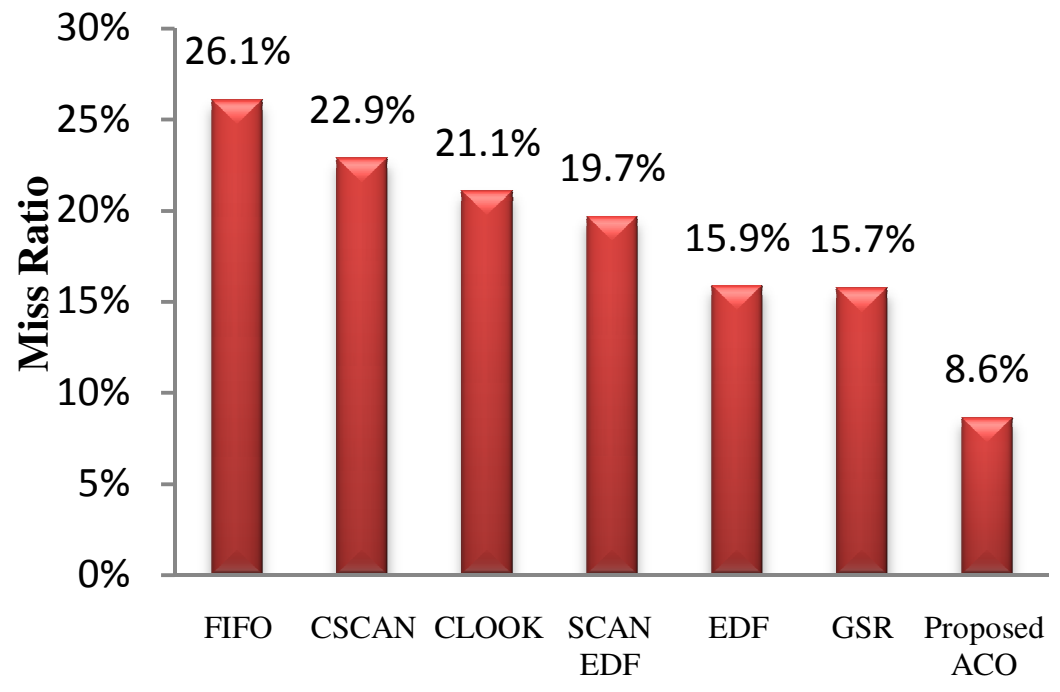

Fig 1- Average miss ratio of problem s in TC1

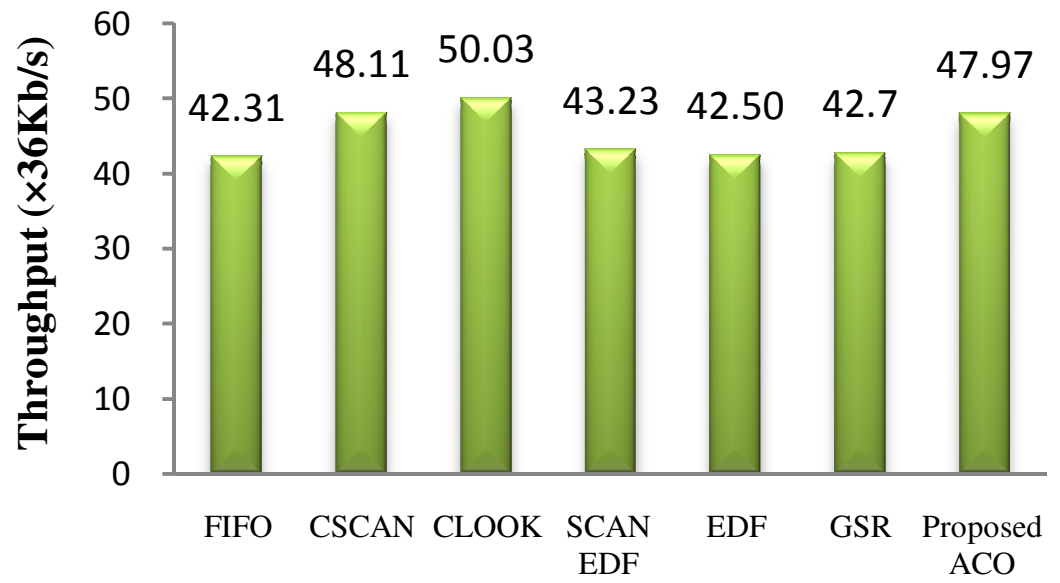

Fig 2- Average throughput of problems in TC1

Figure 3 and 4 show the miss ratio and throughput for all algorithms when they were applied on TC2 problems. Figure 3 shows the average miss ratio over these problems. The values have been calculated in the same manner with figure 1. The proposed method has improved the miss ratio about 6% in the average case in compare with GSR which has the lowest miss ratio among others. Also, our method worked better than FIFO, EDF, GSR and SCAN-EDF in terms of throughput. Its throughput was about 3.6*36*KBps* more than SCAN-EDF that had best throughput among others.

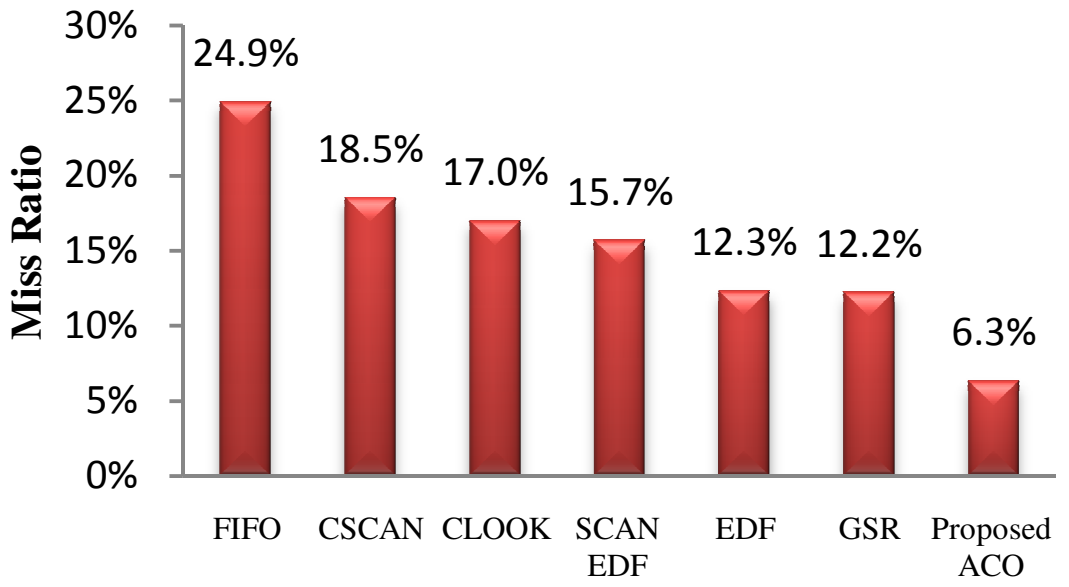

Fig 3- Average miss ratio of problems in TC2

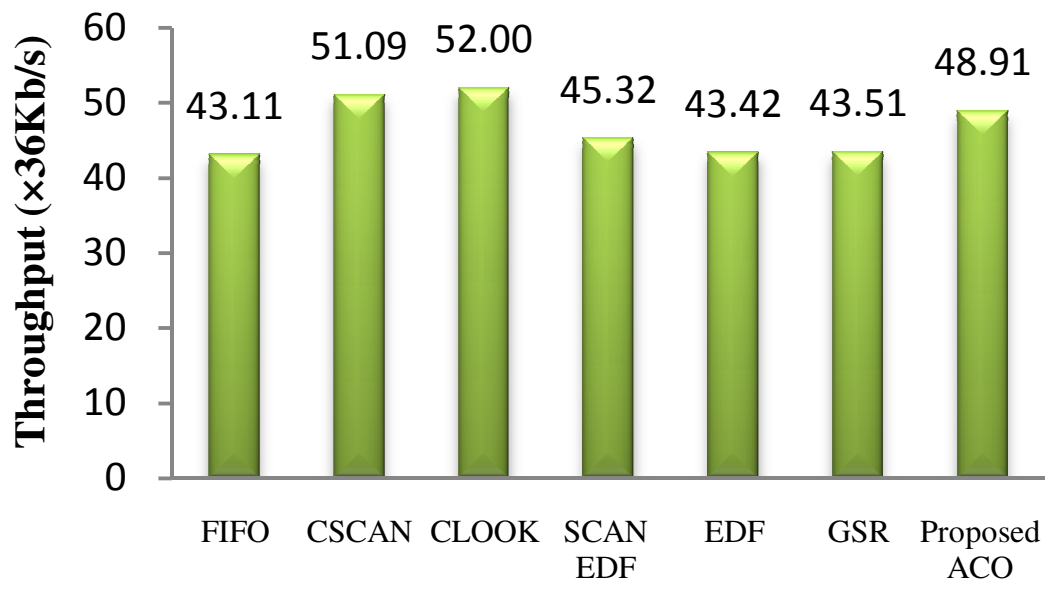

Fig 4- Average throughput of problems in TC2

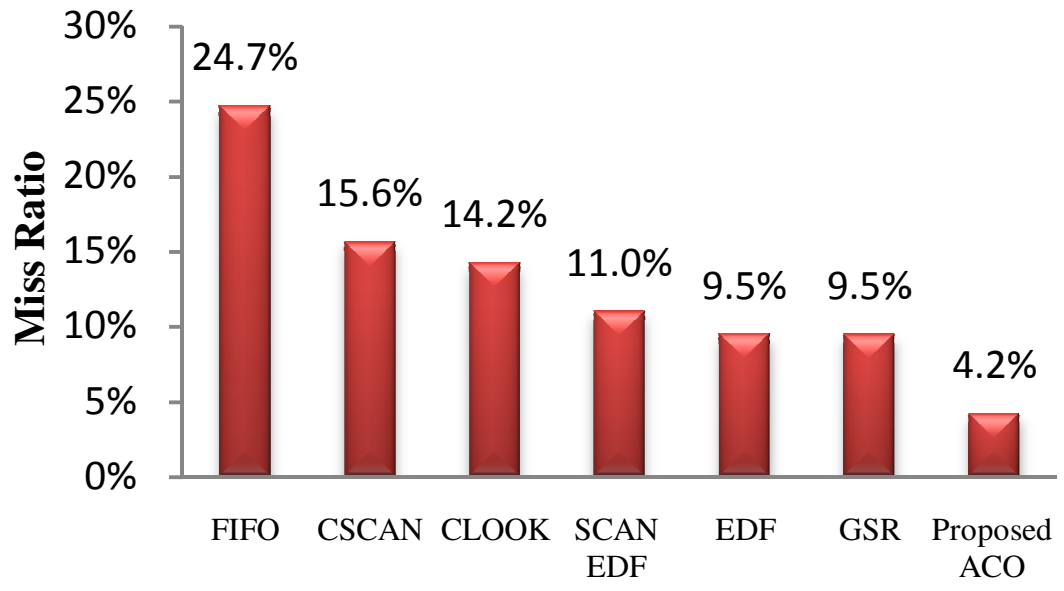

Fig 5- Average miss ratio of problems in TC3

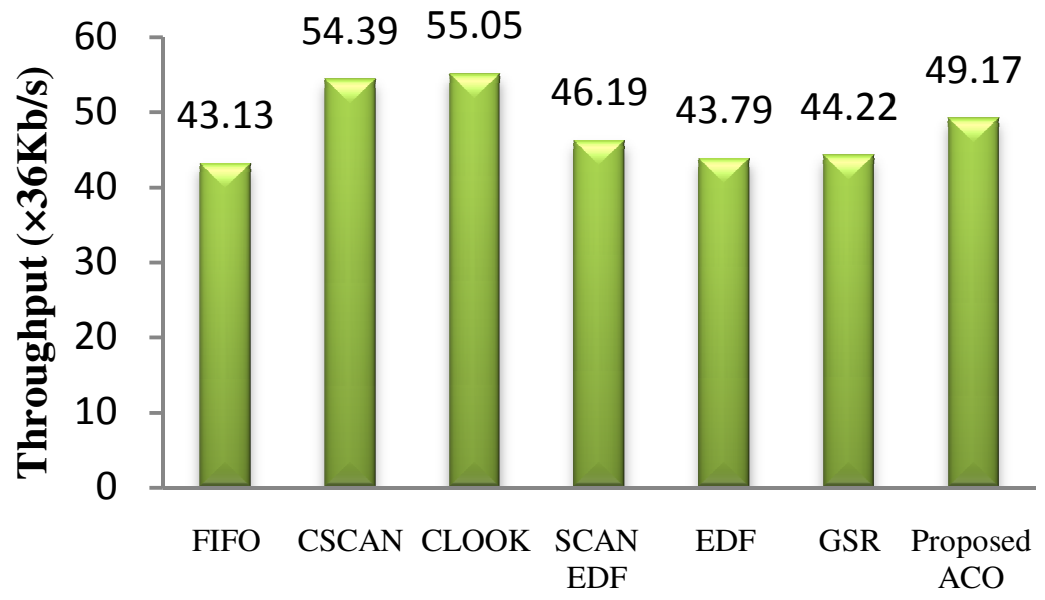

Fig 6- Average throughput of problems in TC3

Figure 5 and 6 shows the average miss ratio and throughput for all algorithms when they were applied on TC3 problems. The proposed ACO has improved the miss ratio about 5% in the average case in compare with GSR which has the lowest miss ratio among traditional methods. It is obvious in the figure 6 that the proposed ACO worked better than FIFO, EDF, GSR and SCAN-EDF in terms of throughput. The throughput is 4.95*36*KBps* more than best throughput among others.

One of the main concerns of proposed method is it running time, because the proposed algorithm is used in a real time environment, high running time may cause some problems in scheduling of other tasks. Fig. 7 shows the average running time of proposed method when it is applied on a queue with different number of tasks. As it seems in the figure 7, the needed time to schedule 20 tasks is about 9*ms*; therefore the scheduling algorithm can run simultaneously with typical tasks. Also, the running time of algorithm increases by increasing number of tasks but its increment is near linear.

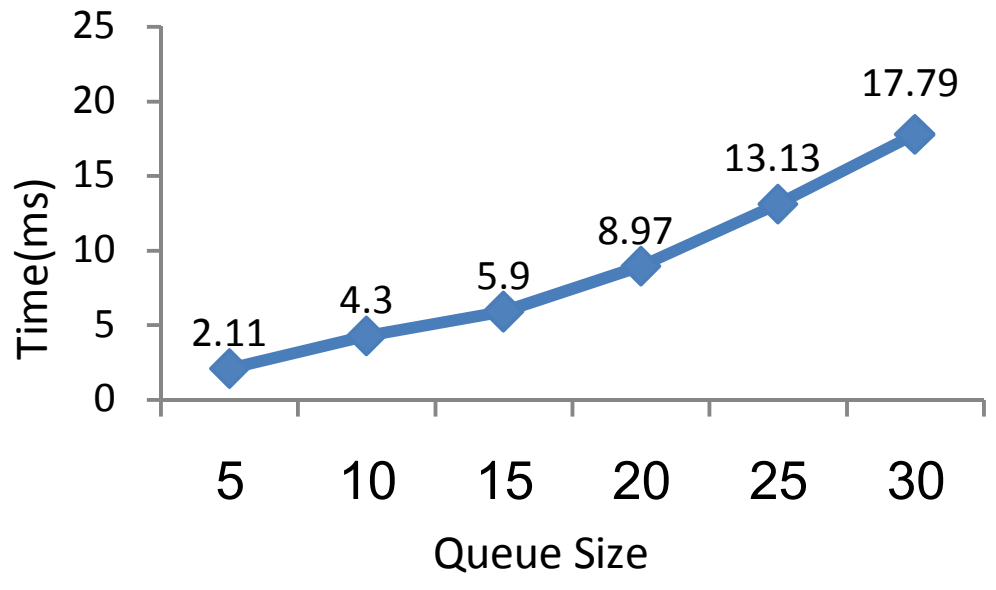

Fig. 7- The average running time of proposed method vs. queue size

Therefore, the proposed method is useful when number of tasks in queue is fair; otherwise it may affect the number of misses. If we are going to use the proposed method for large number of tasks in queue, it should be implemented by hardware [18].

## 5. CONCLUSION

In this paper, a new approach based on Ant Colony Algorithm (ACO) was proposed to solve disk scheduling problem. The simulation results showed that the proposed method has less number of missed tasks versus other related work, and it improved the system throughput at least 5%.

The running time of proposed method is not high when it was implemented with C language; therefore, it can be used to schedule systems that have fair number of tasks. In other cases, a hardware implemented algorithm would be used. In future, we are going to implement this algorithm on FPGA.

## 6. REFERENCES

[1] J. Wilkes, C. Ruemmler, "An introduction to disc drive modeling," *IEEE Computer*, vol. 27, no. 3, pp. 17-29, March 1994.

[2] V. Goebel, P. Halvorsen, O. Anshus, T. Plagemann, "Operating system support for multimedia systems," *The Computer Communications Journal*, vol. 23, no. 3, p. 267–289, 2000.

[3] G. Lipari, J. Santos, R. Santos, "Improving the schedulability of soft real-time open dynamic systems: The inheritor is actually a debtor," *The Journal of Systems and Software*, vol. 81, no. 7, pp. 1093-1104, 2008.

[4] S. Seshadri, Jayant R. Haritsa, S. Thomas,

"Integrating standard transactions in firm real-time database systems," *Information Systems* , vol. 21, no. 1, pp. 3-28, 1996.

[5] G. G. Belford, O. Ulusoy, "Real-time transaction scheduling in database systems," *Information Systems*, vol. 18, no. 9, p. 559–580, 1993.

[6] W.P. Yang, R.C.T. Lee, T.S. Chen, "Amortized analysis of some disk scheduling algorithms: SSTF, SCAN, and N-Step SCAN," *BIT*, vol. 32, no. 4, p. 546–558, 1992.

[7] M. Hofri, "Disk Scheduling: FCFS vs. SSTF Revisited," *Communications of the ACM*, vol. 23, no. 11, p. 645–653, November 1980.

[8] J. A. Stankovic, J. F. Kurose, D. Towsley, S. Chen, "Performance Evaluation of Two New Disk Scheduling Algorithms for Real-Time Systems," *Journal of Real-Time Systems*, vol. 3, no. 3, p. 307–336, 1991.

[9] G. R. Ganger, Y. N. Patt, B. L. Worthington, "Scheduling Algorithms for Modern Disk Drives," in *Proceedings of ACM SIGMETRICS Conference*, May 1994, p. 241–251.

[10] J. Wyllie, K. B. R. Wijayaratne, A. L. N. Reddy, "Disk scheduling in a multimedia I/O system," *ACM Transactions on Multimedia Computing, Communications, and Applications (TOMCCAP)*, vol. 1, no. 1, pp. 37-59, February 2005.

[11] R.I. Chang, W.K. Shih, and R.C. Chang, "Deadline-modification-scan with maximum scannable-groups for multimedia real-time disk scheduling," in *Proceedings of the IEEE Real-Time Systems Symposium*, 1998, pp. 40-49.

[12] R. I. Chang, W. K. Shih, R. C. Chang, H. P. Chang, "Reschedulable-Group-Scan Scheme for Mixed Real-Time/Non-Real-Time Disk Scheduling in a Multimedia System," *Journal of Systems and Software* , vol. 59, no. 2, pp. 143-152, 2001.

[13] R. I. Chang, W. K. Shih, R. C. Chang, H. P. Chang, "GSR: A global seek-optimizing real-time disk-scheduling algorithm," *Journal of Systems and Software* , vol. 80, no. 2, pp. 198-215, 2007.

[14] W. C. Lu, C. N. Chou, W. K. Shih, P. C. Huang, "The NP-hardness and the Algorithm for Real-Time Disk-Scheduling in a Multimedia System," in *Proceedings of the 11th IEEE International Conference on Embedded and Real-Time Computing Systems and Applications (RTCSA'05)*, Hong Kong, 2005, pp. 260-265.

[15] D. Karaboga, S. Ökdem, "Otimal Disk Scheduling Based on ANT Colony Optimization Algorithm," *Erciyes Universiy Journal of the Institute of Science and Technology*, vol. 22, pp. 11-19, 2006.

[16] V. Maniezzo, A. Colorni, M. Dorigo, "The Ant system: Optimization by a colony of cooperating agents," *IEEE Transactions on Systems, Man, and Cybernetics-Part B*, vol. 26, no. 1, pp. 29-41, 1996.

[17] H. H. Hoos, T. Stützle, "MAX-MIN Ant System," *Future Generation Computer Systems*, vol. 16, no. 8, pp. 889-914, 2000.

[18] K.So, M. Guntsch, M. Middendorf, O. Diessel, H. ElGindy, H. Schmeck, B. Scheuermann, "FPGA implementation of population-based ant colony optimization," *Applied Soft Computing*, vol. 4, p. 303–322, March 2004.